\documentclass[prb,twocolumn,showpacs,preprintnumbers,amsmath,amssymb]{revtex4}
\usepackage{graphicx}% Include figure files
\usepackage{dcolumn}% Align table columns on decimal point
\usepackage{bm}% bold math
\usepackage{color}
\newcommand{\mapright}[1]{\smash{\mathop{\hbox to 1.0cm{\rightarrowfill}}\limits^{#1}}}

\begin{document}

\preprint{}

\title{Theory of two-dimensional macroscopic quantum tunneling in YBa${}_2$Cu${}_3$O${}_{7-\delta}$ Josephson junctions coupled to an LC circuit}
% Force line breaks with \\

\author{Shiro Kawabata,$^{1,2,3}$ Thilo Bauch,$^{2}$ and Takeo Kato$^{4}$}

\affiliation{
$^1$Nanotechnology Research Institute (NRI), National Institute of Advanced Industrial Science and Technology (AIST), Tsukuba, Ibaraki, 305-8568, Japan
\\
$^2$Department of Microelectronics and Nanoscience (MC2), Chalmers University of Technology, S-41296 G\"oteborg, Sweden
\\
$^3$CREST, Japan Science and Technology Corporation (JST), Kawaguchi, Saitama 332-0012, Japan 
\\
$^4$The Institute for Solid State Physics (ISSP), University of Tokyo, Kashiwa, Chiba, 277-8581, Japan}

\date{\today}

\begin{abstract}
We investigate  classical thermal activation (TA) and macroscopic quantum tunneling (MQT) for a YBa${}_2$Cu${}_3$O${}_{7-\delta}$(YBCO) Josephson junction coupled to an LC circuit theoretically.
Due to the coupling between the junction and the LC circuit, the macroscopic phase dynamics can be described as the escape process of a fictitious particle with an $anisotropic$ mass moving in a {\it two-dimensional} potential. 
We analytically calculate the escape rate including both  
the TA and MQT regime
by taking into account the peculiar dynamical nature of the system. In  
addtion to large suppression
of the MQT rate at zero temperature, we study details of the  
temperature dependece
of the escape rate across a crossover region.
These results are in an excellent agreement with recent experimental  
data for the MQT and TA rate in
a YBCO biepitaxial Josephson junction. Therefore the coupling to the  
LC circuit is essential in
understanding the macroscopic quantum dynamics and the qubit operation  
based on the YBCO
biepitaxial Josephson junctions.
\end{abstract}

% PACS, the Physics and Astronomy
                             % Classification Scheme.
%\keywords{Suggested keywords}%Use showkeys class option if keyword
                              %display desired
\maketitle

%%%%
%%%%
%%%%
%%%%
%%%%
\section{Introduction}
%%%%
%%%%
%%%%
%%%%
%%%%

Macroscopic quantum tunneling (MQT) has become a focus of interest in physics and chemistry because it can provide a signature of quantum behavior in a macroscopic system.~\cite{rf:MQT1,rf:MQT2,rf:Weiss}
Among several works on MQT, Josephson junctions have been intensively studied.~\cite{rf:MQT2,rf:MQT4}
Heretofore experimental tests of MQT were focused on low-$T_c$ superconductor Josephson junctions.
%%
%%%
%%%
%%%
%===================================
\begin{figure}[b]
\begin{center}
\includegraphics[width=6.0cm]{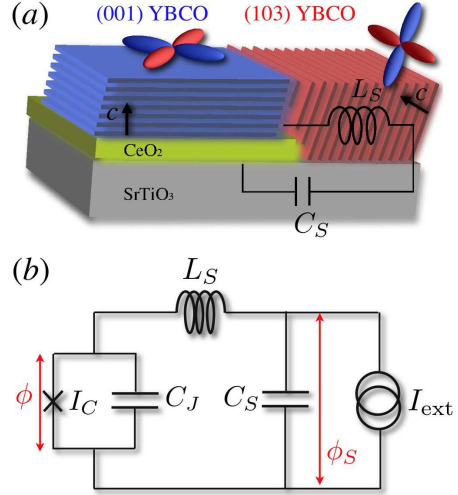}
\caption{(Color online) (a) Schematic of a biepitaxial YBCO junction and (b) the extended circuit model for (a), including the bias current $I_\mathrm{ext}$, the stray capacitance $C_S$ and kinetic inductance $L_S$.
The Josephson junction with the critical current $I_C$ and the capacitance $C_J$ is formed at the boundary between (001) and (103) YBCO electrodes.
$\phi$ and $\phi_S$ are the phase difference across the Josephson junction and the stray capacitance $C_S$, respectively.
}
\end{center}
\label{fig:1}
\end{figure}
%===================================
%%%
%%%
%%%
%%%

Renewed interest in MQT occurred after the recent experimental observations of MQT~\cite{rf:Bauch1,rf:Inomata1,rf:Jin,rf:Matsumoto,rf:Li,rf:Kashiwaya1,rf:YurgensKadowaki} and energy level quantization (ELQ)~\cite{rf:Bauch2,rf:Jin,rf:Inomata2,rf:Kashiwaya2} in high-$T_c$ superconductor Josephson junctions, e.g., YBa${}_2$Cu${}_3$O${}_{7-\delta}$ (YBCO) grain-boundary biepitaxial junctions [see Fig. 1 (a)]~\cite{rf:TafuriKIrtley} and Bi${}_2$Sr${}_2$CaCu${}_2$O${}_{8+\delta}$ intrinsic junctions.~\cite{rf:Yurgens} 
High-$T_c$ Josephson junctions, characterized by a high Josephson plasma frequency $\omega_p$ (up to several THz), exhibit a crossover from the thermal activation (TA) to MQT at relatively high temperatures in comparison with low-$T_c$ systems.

These intriguing findings have definitively opened up the way to quantum systems based on the $d$-wave symmetry of high-$T_c$ superconductors.
The $d$-wave order parameter can be used to create naturally degenerate two-level systems which can offer significant advantages for quantum computation.~\cite{rf:Ioffe,rf:Blais,rf:Blatter,rf:Fominov,rf:Amin,rf:Kato}
However, one of the main arguments against high-$T_c$  qubits based on systems with $d$-wave order parameter symmetry was the presence of low energy excitations, i.e., nodal quasiparticles and zero-energy Andreev bound states (ZESs), destroying quantum coherence.
Recent theoretical works suggest that nodal quasiparticles and ZESs, respectively, give super-Ohmic and Ohmic dissipation on MQT.~\cite{rf:Kawabata1,rf:Kawabata2,rf:Kawabata3,rf:Kawabata4,rf:Kawabata5,rf:Yokoyama,rf:Umeki}

The YBCO biepitaxial grain-boundary junctions which were used in MQT and ELQ experiments,~\cite{rf:Bauch1,rf:Bauch2} have quite novel structure  and several fascinating advantages for the coherent property over other types of high-$T_c$ junctions.
First, the relative orientation of the  $d$-wave order parameter between two YBCO electrodes can be designed artificially.~\cite{rf:Tafuri,rf:Lombardi0}
Therefore it is possible to make a junction in which a lobe of the $d$-wave order parameter of one electrodes is facing a node in the other electrodes as shown in Fig. 1(a).
In such configuration, super-Ohmic dissipation resulting from the nodal quasiparticles can be suppressed.~\cite{rf:Kawabata1,rf:Kawabata4,rf:Yokoyama}
Moreover the junction has an additional tilting of one electrode [(103) YBCO] with respect to (001) YBCO electrodes [see Fig. 1(a)]. 
In such tilt-junctions, the formation of ZESs causing Ohmic dissipation can be inhibited as was predicted by Golubov and Tafuri.~\cite{rf:Golubov}

Recently, it was found from ELQ experiment that the measured bias-current $I_\mathrm{ext}$ dependence of the resonant frequency deviates significantly from what one expects from the Josephson plasma frequency $\omega_p$ in a conventional single Josephson junction.~\cite{rf:Bauch2}
The resonant frequency  is a factor of five less than what one would expect from the estimated values of the Josephson critical current $I_C$ and  the junction capacitance $C_J$.
This fact could be explained  by taking into account the existence of a large kinetic inductance and stray capacitance coupled to the junction.~\cite{rf:Bauch2}

In the biepitaxial junctions, the film  of the (103) YBCO electrode is oriented in such a way that the electric transport has a large component in the $c$-axis direction [see Fig. 1(a)]. 
So the kinetic inductance $L_S$ of the (103) electrode can be significant and is much larger than the Josephson inductance of the junction $L_J$.~\cite{rf:Bauch2} 
Another important effect which was observed in ELQ experiments is the influence of the stray capacitance  $C_S$ of the SrTiO${}_3$ (STO) substrate with a huge dielectric constant [see Fig. 1(a)].
Hence a large part of the measured capacitance is not due to the junction interface but to a distributed stray capacitance $C_S$ which in effect shunts the junction.~\cite{rf:Bauch2}

The influence of $L_S$ and $C_S$ on the macroscopic dynamics is inevitable in the biepitaxial junctions and can be taken into account by an extended circuit model [see Fig.1(b)].~\cite{rf:Bauch2,rf:Lombardi,rf:Rotoli}
In Fig. 1 (b), $\phi$ is the phase difference across the Josephson junction and $\phi_S= (2 \pi/\Phi_0) I_S L_S + \phi$ is the phase difference across the capacitor $C_S$, where $I_S$ is the current through the inductor $L_S$.
As will be mentioned later, the addition of the LC circuit results in a two-dimensional (2D) potential $U(\phi,\phi_S)$ and an anisotropic mass which make the dynamics much more complex than for an ordinary single junction.
In Refs. [\onlinecite{rf:Bauch2,rf:Lombardi,rf:Rotoli}] we have theoretically investigated the $classical$ resonant activation based on the extended circuit model, and found that the bias-current $I_\mathrm{ext}$ dependence of the resonant frequencies  can be quantitatively explained by the normal modes (upper and lower resonant mode) of the system. 
Those normal modes, under certain conditions, can be  completely different from the bare Josephson plasma frequency $\omega_p$.
Therefore the extended circuit model well describes the ELQ experiment in the YBCO biepitaxial  junctions.

In contrast to ELQ experiment, it was experimentally confirmed that the thermal escape process above the crossover temperature $T_\mathrm{co}$ can be quantitatively explained by the single junction model $without$ the LC circuit, $i. e.$, one-dimensional (1D) model.~\cite{rf:Bauch1}
Therefore a natural question to ask is {\it can the extended circuit model explain both the TA and MQT escape dynamics in the YBCO biepitaxial junction consistently?}
In this paper, in order to answer the important open question, we explore  the validity of the extended circuit model for both the TA and MQT escape processes by extending the previous classical theory.~\cite{rf:Rotoli,rf:Lombardi}
Then we will  show that the presence of the LC circuit has negligible influence on the TA escape rate, but  the MQT escape rate is suppressed  considerably due to the coupling to the LC circuit.
This behavior is nicely consistent with a recent experimental result of the temperature dependence of the TA and MQT escape rate.~\cite{rf:Bauch1}
Therefore, {\it the presence of the LC circuit is fundamental and essential in understanding the macroscopic phase dynamics in the YBCO biepitaxial junctions.}

This paper is organized as follows.
In Sec. II, we show the Lagrangian describing the YBCO Josephson junction coupled to the LC circuit.
We also discuss the anisotropy of the mass and the two-dimensional potential profile for this model.
In Sec. III, we derive an effective action from the Lagrangian and show that the system can be mapped into a one-dimensional model.
The crossover temperature $T_\mathrm{co}$, the TA, and the MQT escape rate are calculated in Secs. IV, V, and VI, respectively.
We try to compare theoretical results with recent experimental data of the YBCO biepitaxial Josephson junction in Sec. VII.
Finally, we summarize our results and draw future directions in Sec. VIII.

%%%%
%%%%
%%%%
%%%%
%%%%
\section{Model}
%%%%
%%%%
%%%%
%%%%
%%%%
In this section we derive the Lagrangian for  the Josephson junction coupled to the LC circuit and discuss the anisotropy of the mass and two-dimensional potential structure of this model.
The Hamiltonian of the circuit [Fig. 1(b)] can be written as
\begin{eqnarray}
{\cal H}
&=&
\frac{Q_J^2}{2 C_J}
+
\frac{Q_S^2}{2 C_S}
-E_J \cos \phi
+
\left( \frac{\Phi_0}{ 2 \pi}\right)^2 
\frac{\left( \phi-\phi_S\right)^2}{2 L_S} 
\nonumber\\
&-&
\left( \frac{\Phi_0}{ 2 \pi} \right)
 \phi_S I_\mathrm{ext} 
 ,
\label{eq:1-1}
\end{eqnarray}
where  $Q_{J}=C_{J}(\Phi_0/2 \pi) (d \phi /d t)$ and  $Q_{S}=C_{S} (\Phi_0/2 \pi) $ $(d \phi_S /d t)$ are the charge on the junction and the stray capacitor $C_S$, respectively, $E_J=(\hbar/2 e) I_C$ is the Josephson coupling energy, and $\Phi_0=h/2e$ is the flux quantum.

The Lagrangian is then given by 
\begin{eqnarray}
{\cal L}
\! = 
\frac{C_J}{2}
\left( 
\frac{\Phi_0}{2 \pi}
\frac{\partial \phi}{\partial t}
\right)^2
+
\frac{C_S}{2}
\left( 
\frac{\Phi_0}{2 \pi}
\frac{\partial \phi_S}{\partial t}
\right)^2
-
U(\phi,\phi_S)
,
\label{eq:1-2}
\end{eqnarray}
\begin{eqnarray}
U(\phi,\phi_S)
&=&
E_J \left[
- \cos \phi + \frac{ \left(\phi-\phi_S \right)^2}{2 \eta}
-
\gamma \phi_S
\right]
,
\label{eq:1-3}
\end{eqnarray}
where $\gamma=I_\mathrm{ext}/I_C$ and $\eta \equiv 2 \pi I_C L_S / \Phi_0= L_S/L_{J0}$ with $L_{J0} = \Phi_0/2 \pi I_C$ being the zero-bias Josephson inductance.
The Lagrangian ${\cal L}$ describes the quantum dynamics of a fictitious particle with an anisotropic mass moving in a two-dimensional tilted washboard potential $U(\phi,\phi_S)$.
Therefore the escape paths traverse a two-dimensional landscape, so the macroscopic dynamics in this model becomes more complicated than that in the simple one-dimensional model.

%%
%%%
%%%
%%%
%===================================
\begin{figure}[b]
\begin{center}
\includegraphics[width=8.0cm]{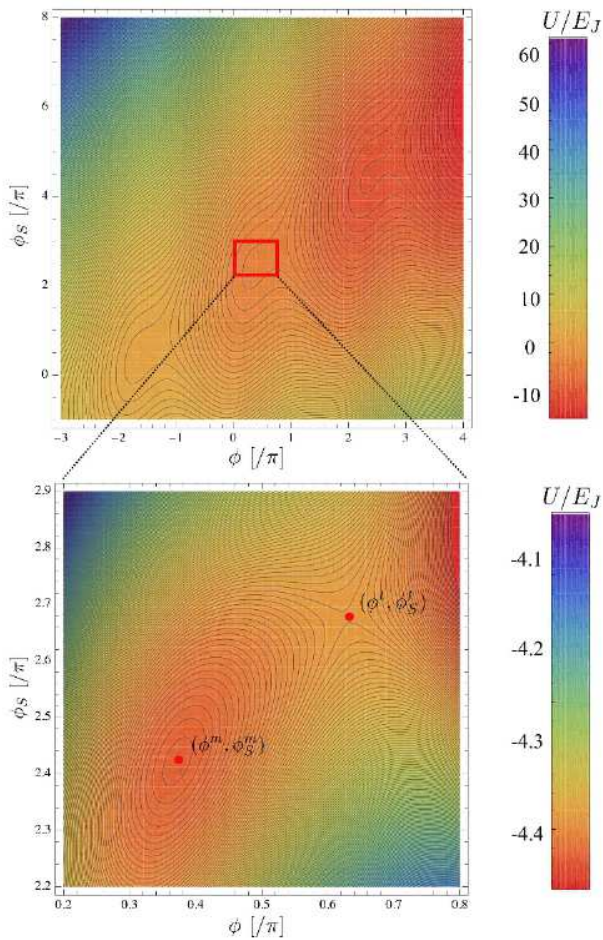}
\caption{(Color online) The two-dimensional potential profile $U(\phi,\phi_S)$ of a YBCO Josephson junction coupled to a LC circuit for $\gamma=I_\mathrm{ext}/I_C=0.92$ and $\eta=L_S/L_{J0}=7$.
Initially the junction oscillate around the local minima at $(\phi^m,\phi_S^m)$.
If $\gamma$ is increased, the junction eventually switches to the finite voltage state by escaping either thermally or via MQT through the saddle point at $(\phi^t,\phi_S^t)$.
}
\end{center}
\label{fig:2}
\end{figure}
%===================================
%%%
%%%
%%%
%%%

The two-dimensional potential profile $U(\phi,\phi_S)$ is shown in Fig. 2.
The mean slope along the $\phi_S$ direction is proportional to the bias current $I_\mathrm{ext}$, and the mean curvature perpendicular to the diagonal direction ($\phi=\phi_S)$ is due to the inductive coupling between the Josephson junction and capacitance $C_S$ characterized by $\eta$.
The local minimum point $(\phi^m,\phi_S^m)$ and the saddle point $(\phi^t,\phi_S^t)$ is given from Eq. (\ref{eq:1-3}) as 
\begin{eqnarray}
\left( \phi^m,\phi_S^m \right)&=&
\left( \sin^{-1} \gamma, \eta \gamma +   \sin^{-1} \gamma \right)
,
\label{eq:1-4}
\\
\left(\phi^t,\phi_S^t \right)&=&
\left( \pi - \sin^{-1} \gamma,  \pi  -   \sin^{-1} \gamma + \eta \gamma \right)
,
\label{eq:1-5}
\end{eqnarray}
Then the potential barrier height $V_0^\mathrm{2D}$ is given by
\begin{eqnarray}
V_0^\mathrm{2D} &\equiv& U \left(\phi^t,\phi_S^t \right) - U \left( \phi^m,\phi_S^m \right) 
\nonumber\\
&=&
E_J \left(
2 \gamma \sin^{-1} \gamma - \pi \gamma + 2 \sqrt{1-\gamma^2}
\right)
,
\label{eq:1-6}
\end{eqnarray}
and is a decreasing function of $\gamma$.
Importantly, the potential barrier height $V_0^\mathrm{2D}$ does not depend on the LC circuit parameters $L_S$ and $C_S$ and the expression of $V_0^\mathrm{2D}$ is identical with the barrier height for usual single Josephson junctions, i.e., one-dimensional model.~\cite{rf:Rotoli}

On the other hand, the overall potential shape can be controlled drastically by changing $L_S$.
In the case of small $L_S$ or equivalently large coupling strength $\eta^{-1}$, the side well becomes shallower and  the potential is almost confined  to the diagonal direction ($\phi=\phi_S$).
On the other hand, when the coupling strength $\eta^{-1}$is decreased, corresponding to a large $L_S$, the confinement to the diagonal direction tends to be weak.
This makes  the escape from a metastable well much easier.
In Secs. V and VI, we quantitatively investigate the $C_S$ and $L_S$ dependence of the thermal and quantum escape rate.

Next we derive an approximate expression of $U(\phi,\phi_S)$ for $\gamma \lesssim 1$. 
By introducing the new coordinate $(x,y) = (\phi - \phi^m, \phi_S - \phi_S^m)$ and assuming $\gamma \lesssim 1$, we can rewrite the Lagrangian as 
\begin{eqnarray}
{\cal L}
&=&
\frac{M}{2}
\left(
\frac{\partial x}{\partial t}
\right)^2
+
\frac{m}{2}
\left(
\frac{\partial y}{\partial t}
\right)^2
-
U(x,y)
,
\label{eq:1-7}
\\
U(x,y)
&=&
E_J \left[
-\frac{x^3}{6}
+
\frac{\sqrt{1- \gamma}}{\sqrt{2}}  x^2
+
\frac{(x-y)^2}{2 \eta}
\right]
\nonumber\\
&=&
U_\mathrm{1D} (x)
+
E_J \frac{(x-y)^2}{2 \eta}
,
\label{eq:1-8}
\end{eqnarray}
where $M=C_J (\Phi_0 / 2 \pi)^2 $ and $m= C_S (\Phi_0 / 2 \pi)^2$.
In this equation, $U_\mathrm{1D} (x)= E_J  \left[ -x^3/6 + \sqrt{(1-\gamma)/2} x^2 \right] \approx (1/2) M \omega_p^2   (x^2 -x^3/x_1)$ is the potential of the Josephson junction without the LC circuit, where $x_1=3\sqrt{1- \gamma^2}$ is the second zero of $U_\mathrm{1D} (x)$ and  $\omega_p = \omega_{p0} (1- \gamma^2)^{1/4}$ is the Josephson plasma frequency with $\omega_{p0} = \sqrt{2 \pi I_C / \Phi_0 C_J}=1/\sqrt{L_{J0} C_J}$ being the zero-bias plasma frequency.

%%%%
%%%%
%%%%
%%%%
%%%%
\section{Derivation of Effective Action: Mapping to a one-dimensional model}
%%%%
%%%%
%%%%
%%%%
%%%%
Here we derive an effective action from the Lagrangian Eq. (\ref{eq:1-7}) of the extended circuit model.
 Note that, in the ELQ experiment, the quality factor $Q$ of the junction was found to be sufficiently large ($Q \approx 40$),~\cite{rf:Bauch2} so we can safely neglect the damping effect resulting from external environments other than the LC circuit.

By using the functional integral method,~\cite{rf:Weiss,rf:Zaikin} the partition function ${\cal Z}$ of the system can be written as
\begin{eqnarray}
{\cal Z} 
=
\int  {\cal D} x (\tau) 
\int  {\cal D} y (\tau) 
\exp
\left(
-\frac{1}{\hbar} \int_0^{\hbar \beta} d \tau {\cal L} [x,y] 
\right)
,
\label{eq:2-1}
\end{eqnarray}
where the functional integral is performed over all the periodic paths with the period $\hbar \beta$.
In this equation, 
$
 {\cal L} [x,y] 
=
(M/2) \dot{x}^2  +(m/2) \dot{y}^2  + U(x,y)
$
 is the Euclidean Lagrangian and $\beta=1/ k_B T$.
The Lagrangian is a quadratic function of $y$ and the coupling term between $x$ and $y$ is linear, so the functional integral over variable $y$ can be performed explicitly by use of the Feynman-Vernon influence functional technique.~\cite{rf:Weiss}
Then the partition function is reduced to a single functional integral over $x$, i.e., ${\cal Z} = \int {\cal D} x(\tau) \exp (- {\cal S}_\mathrm{eff}^\mathrm{2D} [x] /\hbar)$, where the effective action is given by ${\cal S}_\mathrm{eff}^\mathrm{2D}[x]={\cal S}_\mathrm{eff}^\mathrm{1D}[x]+{\cal S}_\mathrm{eff}^\mathrm{ret}[x]$ with~\cite{rf:Chen,rf:Trees}
\begin{eqnarray}
{\cal S}_\mathrm{eff}^\mathrm{1D}[x]
&=&
\int_0^{\hbar \beta}
d \tau \left[
\frac{1}{2} M \dot{x}^2 + U_\mathrm{1D}  (x)
\right] 
,
\label{eq:2-2}
\\
{\cal S}_\mathrm{eff}^\mathrm{ret}[x]
&=&
\frac{1}{4}
\int_{0}^{\hbar \beta} d \tau
\int_0^{\hbar \beta} d \tau'
\left[ x(\tau) - x(\tau') \right]^2  K (\tau-\tau') 
\!
.
\nonumber\\
\label{eq:2-3}
\end{eqnarray}
Thus the dynamics of the phase difference with an anisotropic mass moving in a two-dimensional potential $U(x,y)$ can be mapped into  a simple one-dimensional model.
Note that due to the coupling between the junction and the LC circuit, the effective action ${\cal S}_\mathrm{eff}^\mathrm{2D}[x]$ contains a kind of $dissipation$ action ${\cal S}_\mathrm{eff}^\mathrm{ret}[x]$ in a sense that a retardation (or nonlocal) effect exists.

The nonlocal kernel $K(\tau)$ in Eq. (\ref{eq:2-3}) is defined by
\begin{eqnarray}
K(\tau)
&=&
\frac{1}{2} m \omega_{LC}^3
\frac{
      \cosh \left[  \omega_{LC} \left(  \frac{\hbar \beta}{2} - \left| \tau \right| \right)\right]
      }
      {
      \sinh \left[  \frac{ \hbar \beta \omega_{LC} }{2}  \right]
      }
      \nonumber\\
     &=&
      \frac{M}{\hbar \beta}
       \sum_{n=-\infty}^\infty
       \zeta_n e^{ - i \omega_n \tau}
       ,
       \label{eq:2-4}
\end{eqnarray}
where $\omega_\mathrm{LC}=1/\sqrt{L_S C_S}$ is the LC resonant frequency, $\omega_n= 2 \pi n /\hbar \beta$ is the Matsubara frequency, and $\zeta_n=(C_S/C_J) \omega_{LC}^4 / (\omega_{n}^2 + \omega_{LC}^2)$ is the Fourier coefficient of $K(\tau)$.
The nonlocal kernel $K(\tau)$ is related to the spectral density $J(\omega)$  in the language of the Caldeira-Leggett theory~\cite{rf:Leggett} via
\begin{eqnarray}
K(\tau)
&=&
\frac{1}{\pi} \int_0^\infty d \omega J (\omega) D_\omega (\tau)
,
\label{eq:2-5}
\end{eqnarray}
where $ D_\omega (\tau)$ is the Matsubara Green's function of a free boson, 
$
D_\omega (\tau)
=
(1/\hbar \beta)  \sum_{n=-\infty}^{\infty}  \exp \left(  i \omega_n \tau \right) $ $\left\{ 2 \omega / (\omega_n^2 + \omega^2)  \right\}.
$
In the Caldeira-Leggett theory, the external bath is modeled by a sum of an $infinite$ number of harmonic oscillators which is capable of destroying MQT.~\cite{rf:Weiss,rf:Leggett}
On the other hand, in our case, the bath can be described by a $single$ harmonic oscillator and then $J(\omega)$ is given  as a delta-function, i.e.,
\begin{eqnarray}
J(\omega)
&=&
\frac{\pi}{2} m \omega_{LC}^3 \ \delta (\omega - \omega_{LC})
.
\label{eq:2-4-2}
\end{eqnarray}

For later convenience, we rewrite the effective action ${\cal S}_\mathrm{eff}^\mathrm{ret}[x]$ in terms of the memory kernel $k(\tau)$,~\cite{rf:Weiss}  i.e.,
\begin{eqnarray}
{\cal S}_\mathrm{eff}^\mathrm{ret}[x]
&=&
\frac{1}{2}
\int_{0}^{\hbar \beta} d \tau
 \int_0^{\hbar \beta} d \tau'
x(\tau)  k (\tau-\tau') x(\tau')  
,
\label{eq:2-6}
\\
k(\tau)
&=&
       m \omega_{LC}^2 \sum_{n=-\infty}^\infty
\delta(\tau- n \hbar \beta)
-K(\tau)
\nonumber\\
     &=&
      \frac{M}{\hbar \beta}
       \sum_{n=-\infty}^\infty
       \xi_n e^{ - i \omega_n \tau}
       ,
       \label{eq:2-7}
\end{eqnarray}
where the Fourier coefficient of the memory kernel $\xi_n$ is related to the Fourier transform of the memory-friction kernel $\gamma(t)$,~\cite{rf:Weiss}  $i.e., $ $\hat{\gamma} (\omega)=\int_{-\infty}^\infty d t \gamma(t) e^{i \omega t}$, via
\begin{eqnarray}
\xi_n
=
| \omega_n| \hat{\gamma}(| \omega_n| )
=
\frac{C_S}{C_J}\frac{ \omega_{n}^2  \omega_{LC}^2}{\omega_{n}^2 + \omega_{LC}^2}
.
\label{eq:2-8}
\end{eqnarray}
In below, by using the derived effective action ${\cal S}_\mathrm{eff}^\mathrm{2D}$, we calculate the crossover temperature and the escape rate,  and then discuss the influence of the anisotropic mass and the two-dimensional nature of the potential profile on the macroscopic phase dynamics in the YBCO junction coupled to the LC circuit.

%%%%
%%%%
%%%%
%%%%
%%%%
\section{Crossover Temperature}
%%%%
%%%%
%%%%
%%%%
%%%%
Based on the normal mode analysis at the barrier top of the potential $U(x,y)$, the crossover temperature between the MQT and TA escape process is given by~\cite{rf:GrabertWeiss} 
\begin{eqnarray}
T_\mathrm{co}^\mathrm{2D}
=
\frac{\hbar \omega_\mathrm{R}}{2 \pi k_B},
\label{eq:3-1}
 \end{eqnarray}
where $\omega_R$ is the renormalized trial frequency due to the coupling to the LC circuit and is a positive root of the equation,
$
\omega_\mathrm{R}^2 + \omega_\mathrm{R} \hat{\gamma} (\omega_\mathrm{R}) = \omega_p^2
.
$
By using Eq. (\ref{eq:2-8}), we get 
\begin{eqnarray}
\omega_\mathrm{R}
&=&
\frac{1}{\sqrt{2}}
\omega_p
\sqrt{
\theta 
+
\sqrt{
 \theta^2
+
4 \frac{\omega_\mathrm{LC}^2}{\omega_p^2}
}
}
\label{eq:3-2-1}\\
&\approx&
\left\{
\begin{array}{cl}
\displaystyle{
   \omega_p \sqrt{1- \frac{C_S}{C_J}
   },
                    } 
 & \quad L_J \gg L_S \  \& \ C_J  \gg C_S  \\
\displaystyle{
   \omega_p \sqrt{1-\frac{L_J}{L_S}
  }
   }&   \quad L_J \ll L_S \  \& \ C_J  \ll C_S
\end{array}
\right.
,
\nonumber\\
\label{eq:3-2}
 \end{eqnarray}
for the adiabatic ( $L_J \gg L_S$ and $C_J  \gg C_S $ ) and the nonadiabatic ($L_J \ll L_S$ and $C_J  \ll C_S $ ) limit, where $ \theta \equiv 1- \left( 1+  C_S / C_J \right) \omega_\mathrm{LC}^2 / \omega_p^2$, and $L_J=L_{J0}/\sqrt{1-\gamma^2}$ is the Josephson inductance.
In the adiabatic (nonadiabatic) limit, the LC resonant frequency $\omega_\mathrm{LC}$ is much larger (smaller) than the Josephson plasma frequency $\omega_p$.
As a result, the LC resonant  mode can (cannot) adiabatically follow the dynamics of the Josephson junctions.
Note that, in the YBCO biepitaxial junction  used in the MQT experiment,~\cite{rf:Bauch1,rf:Bauch2} the nonadiabatic case is approximately realized (see Sec. VII).

By substituting Eq. (\ref{eq:3-2}) into Eq. (\ref{eq:3-1}), we get a simple expression for $T_\mathrm{co}$ as
\begin{eqnarray}
T_\mathrm{co}^\mathrm{2D}
&\approx&
\left\{
\begin{array}{cl}
\displaystyle{
   T_\mathrm{co}^\mathrm{1D} \sqrt{1- \frac{C_S}{C_J}
   },
                    } 
 & \quad L_J \gg L_S \  \& \ C_J  \gg C_S  \\
\displaystyle{
  T_\mathrm{co}^\mathrm{1D} \sqrt{1-\frac{L_J}{L_S}
  }
   }&   \quad L_J \ll L_S \  \& \ C_J  \ll C_S
\end{array}
\right.
,
\nonumber\\
\label{eq:3-3}
\end{eqnarray}
where $T_\mathrm{co}^\mathrm{1D}=\hbar \omega_p / 2 \pi k_B$ is the crossover temperature without coupling to the LC circuit.
Therefore, due to the influence of the LC circuit,  $T_\mathrm{co}^\mathrm{2D}$ becomes smaller than  $T_\mathrm{co}^\mathrm{1D}$.
In the nonadiabatic (adiabatic) case, $T_\mathrm{co}^\mathrm{2D}$ reduces with decreasing the kinetic inductance $L_S$ (increasing the stray capacitance $C_S$).

%%%%
%%%%
%%%%
%%%%
%%%%
\section{Thermal Activation Process}
%%%%
%%%%
%%%%
%%%%
%%%%

The TA escape rate well above the crossover temperature $T_\mathrm{co}^\mathrm{2D}$ is given by~\cite{rf:GrabertWeiss}  
\begin{eqnarray}
\Gamma_\mathrm{TA}^\mathrm{2D}
& =&
\frac{\omega_R}{2 \pi}
c_\mathrm{qm}^\mathrm{2D}
\exp \left(  - \frac{V_0^\mathrm{2D}}{k_B T} \right)
,
\label{eq:4-1}
\\
c_\mathrm{qm}^\mathrm{2D}
& =&
 \prod _{n=1}^\infty 
 \frac{\omega_n^2 + \omega_p^2 + \omega_n \hat{\gamma} (\omega_n) }
 {\omega_n^2 - \omega_p^2 + \omega_n \hat{\gamma} (\omega_n) } 
 ,
\label{eq:4-2}
 \end{eqnarray}
where $c_\mathrm{qm}^\mathrm{2D}$ is the quantum-mechanical enhancement factor resulting form stable fluctuation modes and  $\hat{\gamma}(\omega_n)=(C_S/C_J) |\omega_n| \omega_{LC}^2/(\omega_n^2 +\omega_{LC}^2)$ is the Fourier transform of the memory-friction kernel.
As was shown in Sec. II, the potential barrier height is not changed even in the presence of the LC circuit, i.e., $V_0^\mathrm{2D}=V_0^\mathrm{1D}$.
Therefore the coupling to the LC circuit only modifies the prefactor of $\Gamma_\mathrm{TA}$.
Note that the exponent in $\Gamma_\mathrm{TA}$ for the Josephson junction coupled to the LC circuit was discussed by Fistul.~\cite{rf:Fistul}
 Here we  calculate the TA escape rate $\Gamma_\mathrm{TA}^\mathrm{2D}$ including the prefactor in order to discuss the influence of the LC circuit explicitly and compare with experimental results.
The quantum correction $c_\mathrm{qm}$ in the prefactor can be calculated analytically as
\begin{eqnarray}
c_\mathrm{qm}^\mathrm{2D}
& =&
 \frac{  \sinh(\pi \lambda_+^+) \sinh(\pi \lambda_-^+)}
{\sinh(\pi \lambda_+^-) \sin(\pi \lambda_-^-)}
,
\label{eq:4-3}
 \end{eqnarray}
with
\begin{eqnarray}
\lambda_\pm^+
&=&
 \sqrt{
\frac{ p_+ \pm \sqrt{p_+^2 -4 q^2}}{2}
}
,
\\
\lambda_\pm^-
& =&
 \sqrt{
\frac{   \pm p_- +  \sqrt{p_-^2 +4 q^2}}{2}
}
,
\label{eq:4-3-1}
 \end{eqnarray}
where
$
p_\pm
=
\left( 1+ C_S/C_J \right)
\left( \omega_\mathrm{LC}/\omega_1 \right)^2
\pm
\left( \omega_p / \omega_1 \right)^2
,
$
and 
$
q
=
\omega_p \omega_\mathrm{LC} /\omega_1^2.
$
The effective trial frequency $\omega_\mathrm{R}$ in Eq. (\ref{eq:4-1}) is given by Eq. (\ref{eq:3-2-1}).

In the adiabatic limit ($L_J \gg L_S$ and $C_J  \gg C_S $), the quantum enhancement factor $c_\mathrm{qm}^\mathrm{2D}$ (\ref{eq:4-3}) can be simplified to 
\begin{eqnarray}
c_\mathrm{qm}^\mathrm{2D}
&\approx&
 \frac{\sinh \left(  \frac{\hbar \beta \omega_p }{2}  \sqrt{1-  \frac{C_S}{C_J}} \right)}
          {\sin \left(  \frac{\hbar \beta \omega_p}{2}  \sqrt{1-  \frac{C_S}{C_J}} \right)}
.
\label{eq:4-4}
\end{eqnarray}
On the other hand, in the non-adiabatic limit ($L_J \ll L_S$ and $C_J  \ll C_S $), we get
\begin{eqnarray}
c_\mathrm{qm}^\mathrm{2D}
&\approx&
   \frac{\sinh \left(  \frac{\hbar \beta \omega_\mathrm{LC}}{2 } \sqrt{1 - \frac{L_J}{L_S}} \right) \sinh \left(  \frac{\hbar \beta \omega_p}{2 }  \sqrt{1 + \frac{L_J}{L_S}} \right)}
          {\sinh \left(  \frac{\hbar \beta \omega_\mathrm{LC}}{2 } \sqrt{1 + \frac{L_J}{L_S}} \right) \sin \left(  \frac{\hbar \beta \omega_p}{2 } \sqrt{1 - \frac{L_J}{L_S}}  \right)}
.
\nonumber\\
\label{eq:4-5}
\end{eqnarray}
Note that the nonadiabatic limit is almost realized in the actual YBCO junction~\cite{rf:Bauch1} (see Sec. VII). 
The quantum enhancement factor $c_\mathrm{qm}^\mathrm{2D}$ almost coincides with the result without retardation effects,~\cite{rf:Wolynes} i.e., $c_\mathrm{qm}^\mathrm{2D} \approx c_\mathrm{qm}^\mathrm{1D}=   \sinh \left(  \hbar \beta \omega_p  / 2  \right) /    \sin \left(  \hbar \beta \omega_p / 2  \right)$  for both the adiabatic and non-adiabatic limit (see also Secs. VI. C and VII. B).
From Eqs. (\ref{eq:3-2}), (\ref{eq:4-1}), (\ref{eq:4-4}), and (\ref{eq:4-5}), therefore, we can conclude that  the influence of the coupling to the LC circuit on the thermal activation process is quite weak, so the system behaves as a one-dimensional system well above the crossover temperature $T_\mathrm{co}^\mathrm{2D}$.
This result is qualitatively consistent with the experimental observation in the YBCO biepitaxial junction.~\cite{rf:Bauch1}
In Sec. VII we will numerically compare theoretical results with experimental data in the TA regime.

%%%%
%%%%
%%%%
%%%%
%%%%
\section{Macroscopic Quantum Tunneling Process}
%%%%
%%%%
%%%%
%%%%
%%%%
To obtain the MQT escape rate $\Gamma_\mathrm{MQT}^\mathrm{2D}$ for the Josephson junction coupled to the LC circuit below $T_\mathrm{co}^\mathrm{2D}$, the usual procedure is to apply the so-called Im-$F$ method.~\cite{rf:MQT2,rf:Weiss}
 It is based on the calculation of the free energy $F=-(1/\beta) \ln {\cal Z}$ and thus of the partition function ${\cal Z}$ of an unstable system.
 According to the metastable decay theory,~\cite{rf:Weiss} the MQT escape rate has the form $\Gamma_\mathrm{MQT}^\mathrm{2D} = -(2/\hbar) \mathrm{Im} F$.
 In below we derive a  weak retardation condition in which  ${\cal S}_\mathrm{eff}^\mathrm{ret}$ in the effective action ${\cal S}_\mathrm{eff}^\mathrm{2D}$ can be treated as a small perturbation.
 Then we calculate $\Gamma_\mathrm{MQT}^\mathrm{2D}$ for zero temperature and finite temperature ($0<T<T_\mathrm{co}^\mathrm{2D})$ in a weak retardation limit.

%%%
%%%
%%%
%%%
\subsection{Weak retardation condition}
%%%
%%%
%%%
%%%

Here we derive a condition for the weak retardation or nonlocal effect.
The bounce trajectory $x_B (\tau)$ satisfies the Euler-Lagrange equation obtained from the variation principle $\delta {\cal S}_\mathrm{eff}^\mathrm{2D}[x]=0$.
Taking the variation of the effective action Eqs. (\ref{eq:2-2}) and (\ref{eq:2-3}), we then get
\begin{eqnarray}
M \frac{d^2 x_B (\tau)}{ d \tau^2}
-
\frac{\partial U_\mathrm{1D}} {\partial x_B (\tau)} 
\!
-
\!
\int_{0}^{\hbar \beta} 
\!
\!
d \tau'
 k (\tau-\tau') x_B (\tau')  
=0
.
\label{eq:5-1}
\end{eqnarray}
Introducing the Fourier expansion
$
x_B (\tau) 
=
\sum_{n=-\infty}^\infty
\tilde{x}_n^{(B)}  $ $\exp ( i \omega_n \tau)
$
into Eq. (\ref{eq:5-1}) and using Eq. (\ref{eq:2-7}), we obtain the following equation for the Fourier coefficient $\tilde{x}_n^{(B)} $
\begin{eqnarray}
&&
\!
\!
\!M 
\left[
\left( \omega_n^2 + \omega_p^2 \right) 
\tilde{x}_n^{(B)} 
-
\frac{3}{2}
\frac{\omega_p^2}{x_1}
\sum_{m=-\infty}^\infty
\tilde{x}_m^{(B)} 
\tilde{x}_{n-m}^{(B)}
\right]
\nonumber\\
&+&
M \xi_n \tilde{x}_n^{(B)}
=0
.
\label{eq:5-2}
 \end{eqnarray}
For $\omega_n=\omega_p$, we get 
\begin{eqnarray}
2  M \omega_p^2
\tilde{x}_n^{(B)} 
&-&
\frac{3}{2}
\frac{M \omega_p^2}{x_1}
\sum_{m=-\infty}^\infty
\tilde{x}_m^{(B)} 
\tilde{x}_{n-m}^{(B)}
\nonumber\\
&+&
M \xi_n (\omega_n= \omega_p) \tilde{x}_n^{(B)}
=0
.
\label{eq:5-3}
 \end{eqnarray}
The criterion for weak retardation can be obtained from Eq. (\ref{eq:5-3}) by comparing the retardation term ($M \xi_n (\omega_n= \omega_p) \tilde{x}_n^{(B)}$) with the first term ($2  M \omega_p^2 \tilde{x}_n^{(B)}$).
By defining the parameter~\cite{rf:MQT4}
\begin{eqnarray}
\kappa
&\equiv&
\frac{\xi_n (\omega_n= \omega_p) }{2  \omega_p^2}
=
\frac{1}{2}
\frac{C_S}{C_J}
\frac{ \left(\frac{\omega_{LC} }{\omega_p } \right)^2}{1+\left(\frac{\omega_{LC} }{\omega_p } \right)^2}
\nonumber\\
&\approx&
\left\{
\begin{array}{cl}
\displaystyle{
       \frac{1}{2}
         \frac{C_S}{C_J}
                    },
& \quad L_J \gg L_S \  \& \ C_J  \gg C_S  \\
 \\
\displaystyle{
             \frac{1}{2}
             \frac{L_J}{L_S}
                    }
& \quad L_J \ll L_S \  \& \ C_J  \ll C_S  \\
\end{array}
\right.
,
\label{eq:5-4}
 \end{eqnarray}
the weak retardation limit corresponds to the case of $\kappa \ll 1$.
In this case ${\cal S}_\mathrm{eff}^\mathrm{ret}$ in the effective action can be treated as a small perturbation.
Figure 3 shows the $L_S$ and $C_S$ dependence of the retardation parameter $\kappa$.
In the YBCO junction used in the MQT experiment,~\cite{rf:Bauch1,rf:Bauch2} $C_S/C_J \approx 7.3$ and $L_S/L_J \approx 3.2$ as will be estimated in Sec. VII, so $\kappa \approx 0.15$.
Therefore,  the weak retardation condition is almost satisfied in this case.
%%
%%%
%%%
%%%
%===================================
\begin{figure}[t]
\begin{center}
\includegraphics[width=8.0cm]{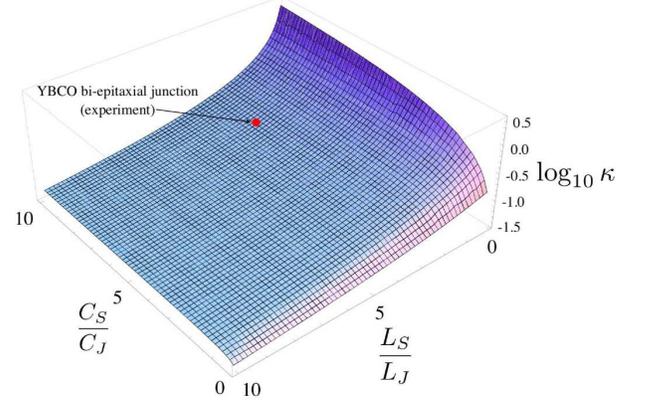}
\caption{
(Color online) $L_S$ and $C_S$ dependence of the parameter $\kappa$ which characterizes the retardation effect due to the junction-LC coupling.
In the YBCO biepitaxial junction which was used in actual MQT experiment, $\kappa \approx 0.15$ (red circle).
}
\end{center}
\label{fig:3}
\end{figure}
%===================================
%%%
%%%
%%%
%%%

%%%
%%%
%%%
%%%
\subsection{Zero Temperature}
%%%
%%%
%%%
%%%

The MQT escape rate at zero temperature is given by $\Gamma_\mathrm{MQT}^\mathrm{2D}= \lim_{\beta \to \infty} (2/ \beta) \mathrm{Im} \ln {\cal Z}$.~\cite{rf:Weiss}
By use of the bounce techniques, the MQT escape rate $\Gamma_\mathrm{MQT}^\mathrm{2D}$ in the weak retardation limit $\kappa \ll 1$ is perturbatively determined by~\cite{rf:Leggett} 
\begin{eqnarray}
\Gamma_\mathrm{MQT}^\mathrm{2D}(T=0)
=
\!
\frac{\omega_p}{2 \pi} \sqrt{120 \pi\left(  B_\mathrm{1D}+ B_\mathrm{ret} \right)  } e^{   - B_\mathrm{1D} - B_\mathrm{ret}  }
,
\label{eq:5-5}
 \end{eqnarray}
where $B_\mathrm{1D} = {\cal S}_\mathrm{eff}^\mathrm{1D} [x_B]/\hbar=36 V_0^\mathrm{1D} / 5 \hbar \omega_p$ and $B_\mathrm{ret}= {\cal S}_\mathrm{eff}^\mathrm{ret} [x_B]/\hbar$ are the bounce exponents, that are the value of the actions evaluated along the bounce trajectory $x_B(\tau)=x_1 \mathrm{sech}^2(\omega_p \tau/2)$.
The bounce action ${\cal S}_\mathrm{eff}^\mathrm{ret}[x_B]$ can be written as 
\begin{eqnarray}
{\cal S}_\mathrm{eff}^\mathrm{ret}[x_B]
&=&
8 \pi^2 
\frac{m \omega_{LC}^2 x_1^2}{\omega_p^4}
\frac{1}{\hbar \beta}
\nonumber\\
&\times&
\sum_{n=-\infty}^\infty \frac{\omega_n^4}
{\left( \omega_n^2 + \omega_{LC}^2 \right) \sinh^2 \left( \frac{\pi \omega_n}{\omega_p} \right)} 
.
\label{eq:5-6}
\end{eqnarray}
In the limit of zero temperature,  we have
%
%
%
%
%\begin{widetext}
\begin{eqnarray}
\!
{\cal S}_\mathrm{eff}^\mathrm{ret}[x_B]
\!
\!
\!
&=&
\!
\!
\!
\frac{
8 \pi
m x_1^2
\omega_{LC}^2
}{\omega_p} 
\int_0^\infty
d z
\frac{z^4}
{\left\{ z^2 + \left( \frac{\omega_{LC}}{\omega_p} \right)^2 \right\} 
\sinh^2 \left( \pi z \right)} 
\nonumber\\
\!
&\approx&
\!
\left\{
\begin{array}{cl}
\displaystyle{\frac{4}{15} m \omega_p  x_1^2} &  \quad L_J \gg L_S \  \& \ C_J  \gg C_S   \\
\displaystyle{\frac{4}{3} m \left( \frac{\omega_{LC}}{\omega_p} \right)^2 \omega_p  x_1^2}&  \quad L_J \ll L_S \  \& \ C_J  \ll C_S
\end{array}
\right.
\!
\!
.
\nonumber\\
\label{eq:5-7}
\end{eqnarray}
%\end{widetext}
%
%
%
Therefore the total bounce exponent is given by 
\begin{eqnarray}
B_\mathrm{1D} + B_\mathrm{diss} 
&=&
\frac{8}{15 \hbar}
(M +\delta M) \omega_p x_1^2
,
\label{eq:5-8}
\end{eqnarray}
where 
\begin{eqnarray}
\frac{\delta M}{M}
&=&
15 \pi \frac{L_J}{L_S}
\int_0^\infty
d z
\frac{z^4}
{\left\{ z^2 + \left( \frac{\omega_{LC}}{\omega_p} \right)^2 \right\} 
\sinh^2 \left( \pi z \right)} 
\nonumber\\
&\approx&
\left\{
\begin{array}{cl}
\displaystyle{\frac{1}{2}  \frac{C_S}{C_J}} &  \quad L_J \gg L_S \  \& \ C_J  \gg C_S \\
\displaystyle{\frac{5}{2}   \frac{L_J}{L_S} }&  \quad L_J \ll L_S \  \& \ C_J  \ll C_S
\end{array}
\right.
,
\label{eq:5-9}
\end{eqnarray}
is the retardation correction to the mass $M$.
Thus, due to the influence of the LC circuit, the bounce exponent is increased with respect to the one-dimensional case.

By substituting Eq. (\ref{eq:5-8}) into Eq. (\ref{eq:5-5}), we finally get the zero-temperature MQT escape rate  for $\kappa \ll 1$ as
\begin{eqnarray}
\Gamma_\mathrm{MQT}^\mathrm{2D}  (T=0)
&=&
\frac{\omega_p}{2 \pi} \sqrt{864 \pi \frac{V_0^\mathrm{1D}}{\hbar \omega_p}  \left( 1+\frac{\delta M}{M} \right) }
\nonumber\\
&\times&
 \exp \left[ 
 - \frac{36}{5} \frac{V_0^\mathrm{1D}}{\hbar \omega_p}  \left( 1+\frac{\delta M}{M} \right) 
 \right]
 .
 \label{eq:5-10}
 \end{eqnarray}
Therefore the coupling to the LC circuit effectively increases the barrier height, $i.e.,$ $V_0^\mathrm{1D} \to V_0^\mathrm{1D} (1+\delta M/M)$. 
This behavior is consistent with the result for low-$T_c$ junctions coupled to an LC circuit which was derived in a different context.~\cite{rf:Esteve,rf:Trees,rf:Martinis}
In contrast to the TA regime, the coupling to the LC circuit reduces the MQT escape rate $\Gamma_\mathrm{MQT}^\mathrm{2D}$ considerably.
Therefore the anisotropic mass and the two-dimensional nature of the potential profile have large influence on the MQT escape process in the low-temperature regime.

\subsection{Finite Temperature}

In this section we calculate the MQT escape rate at finite temperature ($0 < T < T_\mathrm{co}^\mathrm{2D}$) for the weak retardation limit ($\kappa \ll 1$) and show that the $T$  
dependence of the  MQT escape is drastically  influenced  by the coupling to the LC circuit.
At finite-temperature the MQT escape rate is enhanced by thermal fluctuations.
If the temperature is much lower than $T_\mathrm{LC}=\hbar \omega_\mathrm{LC}/ 2 \pi k_B$, the finite temperature enhancement to the $\Gamma_\mathrm{MQT}^\mathrm{2D} (T=0)$ is negligible in contrast  to the system with an Ohmic dissipative environment (see Appendix A).
As will be shown later, however, at sufficiently high temperature but still smaller than $T_\mathrm{co}^\mathrm{2D}$, the coupling to the LC circuit gives a large enhancement to the $\Gamma_\mathrm{MQT}^\mathrm{2D} (T=0)$.

The finite-temperature bounce trajectory $x_B^T (\tau)$ is given by a periodic solution in the inverted potential $-U^\mathrm{1D}(x)$ with energy $-E$ ($0<E< V_0^\mathrm{1D}$).
The corresponding solution is given by
$
x_B^T (\tau) / x_1
=
q_2 + (q_1 - q_2) \mathop{\rm cn^2} (\lambda \omega_p \tau | k)
$,
where $q_3 \le q_2 \le q_1$ are the solutions of the equation $q^2(1-q)=4E/27 V_0^\mathrm{1D}$.~\cite{rf:Zweger}
In this equation $\mathop{\rm cn} $ denotes the Jacobi elliptic function with the parameter $k=\sqrt{(q_1-q_2)/(q_1-q_3)}$ and the coefficient $\lambda=\sqrt{q_1-q_3}/2$.
The period $\hbar \beta$ of the finite-temperature bounce $x_B^T (\tau)$ is given by the complete elliptic integral of the first kind $K(k)=\int_0^1 dx \left [ (1-x^2)(1-k x^2)\right]^{-1/2}$ as 
$
\hbar \beta
=
4 K(k) / \omega_p \sqrt{q_1-q_2} 
$.~\cite{rf:Yasui}
The energy $E$ is related to the temperature $\beta^{-1}$ via this equation.
In the limit of zero-temperature ($E \to 0$), we recover the zero temperature bounce solution $x_B (\tau)=x_1 \mathrm{sech}^2(\omega_p \tau /2 )$.

By substituting $x_B^T (\tau)$ into the effective action, a finite-temperature bounce action ${\cal S}_T^\mathrm{2D}[x_B]$ can be evaluated as
\begin{eqnarray}
{\cal S}_T^\mathrm{2D} [x_B]
&=&
{\cal S}_T^\mathrm{1D} [x_B]+{\cal S}_T^\mathrm{ret}[x_B]
,
 \label{eq:5-11}
 \\
{\cal S}_T^\mathrm{1D} [x_B]
&=&
M \int_0^{\hbar \beta} d \tau {\dot x}_B^T(\tau)^2
+
E \hbar \beta
\nonumber\\
&=&
\frac{\lambda \omega_p M (q_1-q_2)^2}{15 k^2}
F(\lambda \omega_p \hbar  \beta,k)
+
E \hbar \beta
,
 \label{eq:5-12}
\\
{\cal S}_T^\mathrm{ret}[x_B]
&=&
\frac{1}{2}
\int_0^{\hbar \beta} d \tau  
 d \tau'
K(\tau-\tau')
\left[  x_B^T(\tau) - x_B^T(\tau') \right]^2
\nonumber\\
&=&
\frac{1}{2}
m \omega_\mathrm{LC}^2 \hbar \beta
\sum_{n=-\infty}^\infty
\frac{\omega_n^2}{\omega_n^2 + \omega_\mathrm{LC}^2} \left| x_B^T (n)\right|^2
,
 \label{eq:5-13}
\end{eqnarray}
where $x_B^T (n) = (1/ \hbar \beta) \int_0^{\hbar \beta} d \tau  x_B^T(\tau) \exp \left( i \omega_n \tau \right)$ is the Fourier transform of the finite temperature bounce $x_B^T (\tau)$ and the function $F(c,k)$ is defined by 
\begin{eqnarray}
F(c,k)
&=&
\frac{ \mathop{\rm dn} (c | k)}{\sqrt{1- k  \mathop{\rm sn} (c | k)^2}}
\left[ 
    d(k) + e(k)  E \left(   \mathop{\rm am} (c | k), k \right)
\right.
   \nonumber\\ 
   &+&
      \left.
        k  \mathop{\rm sn} (c | k)  \mathop{\rm cn} (c | k)
f(c,k) \sqrt{g(c,k)}
\right]
,
 \label{eq:5-14}
 \end{eqnarray}
with $\mathop{\rm sn}(c|k)$  and $\mathop{\rm dn}(c|k)$  the Jacobi elliptic functions, $\mathop{\rm am}(c|k)$ is the Jacobi amplitude,  $E(\phi,k)=\int_0^\phi d \theta \sqrt{1- k^2 \sin^2 \theta}$ is the elliptic integral of the second kind, $d(k)=-4\left\{  2 + k(k-3) \right\}$, $e(k)= 8 \left\{ 1 + k(k-1) \right\}$, $f(c, k) = k-2-3k   \mathop{\rm cn^2} (c | k)+ 3 k  \mathop{\rm sn^2} (c | k)$, and $g(c,k)=2 k  \mathop{\rm cn^2} (c | k) - 2 \left\{k-2+k   \mathop{\rm sn^2} (c | k) \right\}^2$.

The prefactor of the MQT escape rate $\Gamma_\mathrm{MQT}^\mathrm{2D} (T)$ is determined from the fluctuation modes around the bounce solution $x_B^T (\tau)$ and can be calculated from the Gel'fand-Yaglom formula~\cite{rf:Gerfand,rf:Yasui} as
\begin{eqnarray}
\!
A(T)
\!
&=&
\!
\frac{1}{2}
\sqrt{\frac{9 M \omega_p^3 (1-\gamma^2)}{2 \pi \hbar}}  
\frac{  
        (q_1-q_3)^{3/4}  (q_1-q_2) (1-k^2) 
        }
{
   \sqrt{  a(k)E(k) + b(k) K(k) }
}
\nonumber\\
&\times&
\sinh \left( \frac{\omega_p \beta \hbar}{2} \right)
,
 \label{eq:5-15}
 \end{eqnarray}
where $a(k)=2(k^4-k^2+1)$, $b(k)=(1-k^2)(k^2-2)$, and $E(k)$ is the elliptic integral of the first kind.~\cite{rf:prefactor} 
For $E \to 0$, $A(T)$ is reduced to well-known zero-temperature result, i.e., $A(T=0)=(\omega_p / 2 \pi )\sqrt{864 \pi V_0^\mathrm{1D}/ \hbar \omega_p}$.
Thus we analytically obtain the finite-temperature MQT escaper rate for the weak retardation limit: 
\begin{eqnarray}
\Gamma_\mathrm{MQT}^\mathrm{2D} (T)
&=&
A(T)
\exp 
\left( 
 -\frac{{\cal S}_T^\mathrm{2D} [x_B]}{\hbar}
\right)
.
 \label{eq:5-16}
 \end{eqnarray}
This result is smoothly matched to the zero-temperature expression (\ref{eq:5-10}) when $T \to 0$.

%%
%%%
%%%
%%%
%===================================
\begin{figure}[b]
\begin{center}
\includegraphics[width=7cm]{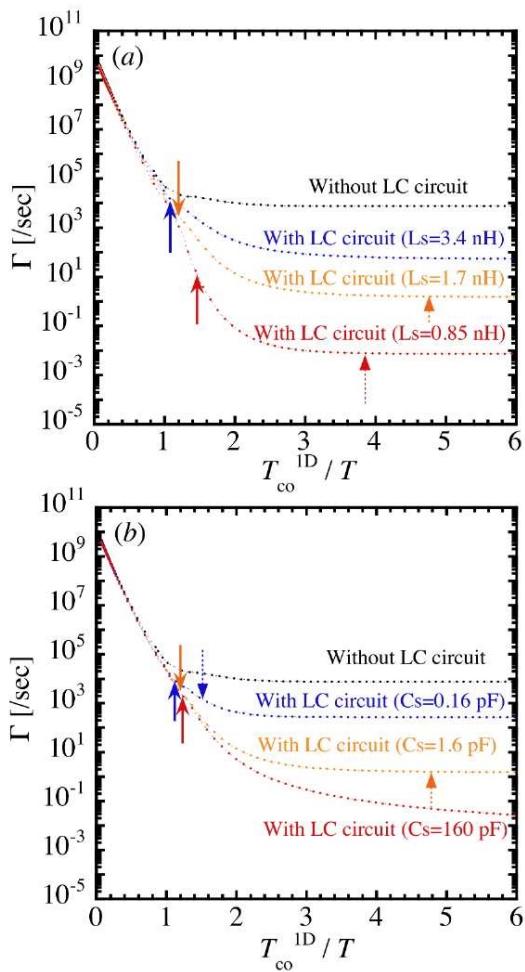}
\caption{(Color online) Arrhenius plot of the escape rate $\Gamma(T)$ for the YBCO junction coupled to the LC circuit.
(a) $L_S$ dependence of $\Gamma(T)$ for $C_S=1.6$ pF. 
 (b)  $C_S$ dependence of $\Gamma(T)$ for $L_S=1.7$ nH.
We choose $\gamma=0.9$, $I_C=1.4$ $\mu$A ($L_J=$ 0.54 nH) and $C_J=0.22$ pF for both graphs.
In the above parameter range, the weak retardation condition ($\kappa \ll 1 $) is satisfied.
Solid and dotted arrows indicate the crossover temperature $T_\mathrm{co}^\mathrm{2D}$ and $T_\mathrm{LC}$, respectively.
}
\end{center}
\label{fig:4}
\end{figure}
%===================================
%%%
%%%
%%%
%%%

The Arrhenius plot of the escape rate $\Gamma_\mathrm{MQT}^\mathrm{2D} (T)$ and $\Gamma_\mathrm{TA}^\mathrm{2D} (T)$  are shown in Fig. 4.
Note that we have used an interpolate crossover-formula~\cite{rf:Weiss} near the crossover temperature $T_\mathrm{co}^\mathrm{2D}$ because $\Gamma_\mathrm{MQT}^\mathrm{2D} (T)$ and $\Gamma_\mathrm{TA}^\mathrm{2D} (T)$ diverges as $T \to T_\mathrm{co}^\mathrm{2D}$ due to the breakdown of the Gaussian approximation (see Appendix B).
In the numerical calculation, we have used $I_C=1.4$ $\mu$A and $C_J=0.22$ pF.
These parameters agree with those directly measured or estimated from experiment,~\cite{rf:Bauch1,rf:Bauch2} allowing a comparison between numerical and experimental data (see Sec. VII).
The TA escape rate $\Gamma_\mathrm{TA}^\mathrm{2D} (T)$ above $T_\mathrm{co}^\mathrm{2D}$ is almost same as in the case without the LC circuit as clearly seen from Fig. 4.
On the other hand, $\Gamma_\mathrm{MQT}^\mathrm{2D} (T)$ is considerably reduced due to the coupling to the LC circuit.
As was explained in Sec. II, the MQT escape rate is significantly reduced with decreasing $L_S$.
This is due to the drastic change in the two-dimensional potential profile.
Similarly the MQT escape rate is decreasing with increasing $C_S$.
This behavior can be explained as follows.
In the case of small $C_S$, the kinetic energy to the $y$ (or $\phi_S)$ direction becomes small.
Therefore, due to the strong anisotropy of the mass, MQT along the external escape direction which connects the minimum  $(\phi^m,\phi_S^m)$ and the saddle point $(\phi^t,\phi_S^t)$ (see Fig. 2) is inhibited.

Interestingly, the temperature dependence of the MQT escape rate $\Gamma_\mathrm{MQT}^\mathrm{2D} (T)$ for the two-dimensional system is quite different from that of the one-dimensional system.
In the case without a LC circuit, $\Gamma_\mathrm{MQT}^\mathrm{1D} (T)$ is almost constant below the crossover temperature $T_\mathrm{co}^\mathrm{1D}$.
On the other hand, in the case of small $L_S$ and large $C_S$, $\Gamma_\mathrm{MQT}^\mathrm{2D} (T)$ significantly depends on the temperature in the range of $T_\mathrm{LC}< T < T_\mathrm{co}^\mathrm{2D}$.
Therefore, from the temperature dependence of the MQT escape rate, we can distinguish qualitatively the anisotropy of the mass and the dimensionality of the quantum phase dynamics.

%%%%
%%%%
%%%%
%%%%
%%%%
\section{Comparison with Experiments}
%%%%
%%%%
%%%%
%%%%
%%%%
In this section, we first summarize the experimental device parameters relevant to the TA and MQT phenomena of a YBCO grain-boundary biepitaxial Josephson junction (Sec. VII. A).
Then, in order to check the validity of the extended circuit model~\cite{rf:Bauch2,rf:Lombardi,rf:Rotoli} for the TA and MQT escape process, we try to compare our result with the experimental data~\cite{rf:Bauch1} of the switching current distribution at the high-temperature TA and the low temperature MQT regimes (Sec. VII. B).

\begin{table}[b]
\begin{center}
\begin{tabular}{lc}
 \hline
 \hline
Parameters & Estimated value \\ 
  \hline
$I_C$: Josephson critical current & 1.4 $\mu$A \\
$L_{J0}$: Josephson inductance & 0.24 nH \\
$L_{J}(\gamma=0.9)$: Josephson inductance & 0.54 nH \\
$C_J$: Capacitance of the junction & 0.22 pF \\
 \hline
 $L_S$: Inductance of the LC circuit & 1.7 nH  \\
 $C_S$: Capacitance of the LC circuit & 1.6 pF \\
 \hline
  \hline
\end{tabular}
\end{center}
\caption
{Experimental junction parameters used for the estimation of the TA and MQT escape rate.}
\label{table1} 
\end{table}
%
%
%

%%%%
%%%%
\subsection{Experimental parameters}
%%%%
%%%%
Experimental parameters used in numerical calculation are given in Table I.
The values given in the table are indeed typical ones in actual experiments.
By the fitting to the Kramers formula with experimental data of the TA escape rate well above the crossover temperature, $I_C$ was experimentally estimated as $I_C=1.4$ $\mu$A which corresponds to $L_{J0}=0.24$ nH.~\cite{rf:Bauch1}
The values of $L_S=1.7$ nH and $C_S=1.6$ pF have been directly determined from the ELQ experiments.~\cite{rf:Bauch2}
Remaining unknown junction parameter is $C_J$.
As will be shown in the next section, this value is estimated as $C_J=0.22$ pF.
This allows for a qualitative comparison between numerical and experimental data of the zero temperature MQT escape rate.
 In this case $C_S/C_J \approx 7.3$ and $L_S / L_{J0} \approx 7.2$ ($L_S/L_{J} \approx 3.2$ for $\gamma=0.9$).
Therefore, in the actual junction, the nonadiabatic limit is approximately realized.

%%%%
%%%%
\subsection{Numerical results}
%%%%
%%%%

In this section, we numerically calculate the switching current distribution $P(\gamma)$ which is related to the escape rate $\Gamma$ as~\cite{rf:Voss,rf:Garg} 
\begin{eqnarray}
P(\gamma)=\frac{1}{v}
 \Gamma (\gamma) \exp
\left[
- \frac{1}{v}
\int_0^{\gamma} \Gamma (\gamma') d \gamma'
\right]
,
 \label{eq:6-1}
\end{eqnarray}
where $v \equiv \left| d \eta / d t \right| $ is the sweep rate of the external bias current.
In the actual experiment,~\cite{rf:Bauch1} the temperature dependence of the full width at half maximum (HMFW) $\sigma$ of $P(\gamma)$ is measured  as shown in Fig. 5.

%%
%%%
%%%
%%%
%===================================
\begin{figure}[b]
\begin{center}
\includegraphics[width=7cm]{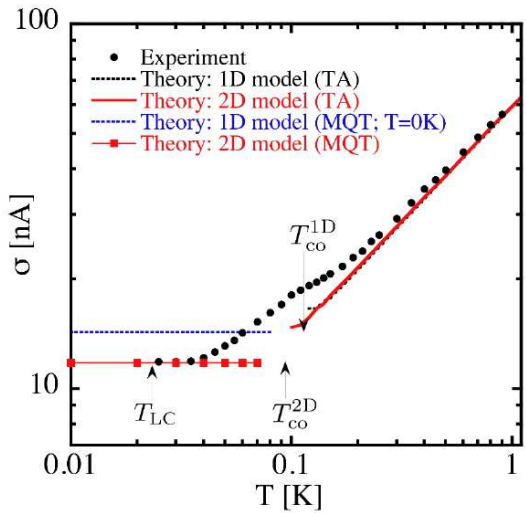}
\caption{(Color online) The temperature dependence of the full width at half maximum $\sigma$ of the switching current distribution $P(\gamma)$.
Both the one- (dotted black line) and two-dimensional models (red solid line) give almost same result above the crossover temperature.
The calculated $\sigma$  from the zero-temperature MQT escape rate for the one-dimensional model and the finite-temperature MQT escape rate for two-dimensional model [Eq. (44)] are shown by dashed-dotted (blue) line and (red) squares, respectively.
Experimental data of $\sigma$ (black circles) for a YBCO biepitaxial Josephson junction~\cite{rf:Bauch1} is also plotted.
We choose $I_C=1.4$ $\mu$A  ($L_{J0}=0.24$ nH), $C_J=0.22$ pF, $L_S=1.7$ nH, $C_S=1.6$ pF and $v I_C=1.0$ mA/s.
Arrows indicate the several characteristic temperatures, i.e., $T_\mathrm{LC}$, $T_\mathrm{co}^\mathrm{1D}$, and $T_\mathrm{co}^\mathrm{2D}$ for $\gamma=0.9$.
}
\end{center}
\label{fig:5}
\end{figure}
%==================================
%%%
%%%
%%%
%%%

First we investigate the TA regime.
In Fig. 5 we show the temperature dependence of $\sigma$ in the TA escape regime (red solid and black dotted lines).
In calculation we have substituted Eqs. (\ref{eq:4-1}) and (\ref{eq:4-3}) into Eq. (\ref{eq:6-1}).
Both the one- and two-dimensional model give good agreement with the experimental data (black circles) well above the crossover temperature ($T_\mathrm{co}^\mathrm{2D} \sim 0.09$K).
Therefore, in the TA regime, the system can be treated as a one-dimensional model without the LC circuit.

In the MQT regime, the measured saturated value of $\sigma$  at $T=0.03$ K is found to be 11.9 nA.~\cite{rf:Bauch1}
From the numerical estimation of $\sigma$, we found that $C_J =0.22$ pF gives good agreement with the experimental value of $\sigma$ as shown in Fig. 5.
The obtained value of $C_J$ is consistent with the estimated value  $C_J \approx 0.16$ pF$\approx 0.1 C_S$ based on the geometry of the junction.~\cite{rf:Rotoli,rf:CJestimation}
Therefore, we can conclude that the extended circuit model quantitatively explain the MQT experiment~\cite{rf:Bauch1} in the YBCO biepitaxial junction near zero temperature.

In order to test the validity of the extended circuit model more systematically, experimental  measurements of HMFW $\sigma$ by changing $L_S$ and $C_S$ are needed.~\cite{rf:Sin2phi}
In Fig. 6 we show $L_S$ and $C_S$ dependence of $\sigma$. 
If we use the substrates with low dielectric constant $\epsilon$, e.g., MgO ($\epsilon \sim 9.6$) and LaAlO${}_3$ ($\epsilon \sim 23$), the reduction in $\sigma$ with respect to the one-dimensional model (blue dotted line in Fig. 6) becomes small.

%%
%%%
%%%
%%%
%===================================
\begin{figure}[tb]
\begin{center}
\includegraphics[width=7cm]{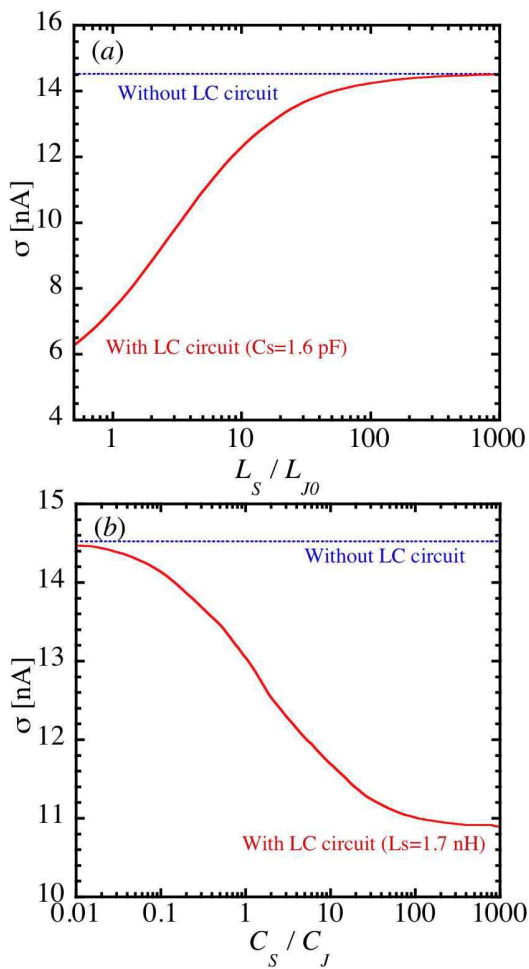}
\caption{(Color online) (a) $L_s$ and (b) $C_s$ dependence of the full width at half maximum $\sigma$ of the switching current distribution $P(\gamma)$ at $T=0$ K for the case with (red solid) and without (blue dotted) the LC circuit.
We choose the parameters as $I_C=1.4$ $\mu$A ($L_{J0}=0.24$ nH), $C_J=0.22$ pF and $v I_C=1.0$ mA/s.
}
\end{center}
\label{fig:6}
\end{figure}
%===================================
%%%
%%%
%%%
%%%

%%%%
%%%%
%%%%
%%%%
%%%%
\section{Conclusions}
%%%%
%%%%
%%%%
%%%%
%%%%

In the present work, the TA and the MQT escape process of the YBCO Josephson junction coupled to the LC circuit has been analyzed by taking into account the anisotropy of the mass and the two-dimensional nature of the phase dynamics.
Based on the Feynman-Vernon approach,  the effective one-dimensional action is derived by integrating out the degree of freedom of the LC circuit.
We found that the coupling to the LC circuit gives negligible reduction for TA escape rate.
On the other hand, we also found that the MQT escape rate is considerably reduced due to the coupling between the junction and  the LC circuit.
More importantly,  the temperature dependence  of the MQT escape rate for the YBCO junction coupled with the LC circuit is  quite different from that without the LC circuit.
These theoretical results are in an excellent agreement with experimental data of the YBCO biepitaxial Josephson junction.~\cite{rf:Bauch1}
Therefore we can conclude that {\it the anisotropy of the mass and the two-dimensional nature of the potential profile due to the coupling to the LC circuit are quite important and essential to understand macroscopic quantum phenomena and qubit operation in such systems.}

There, however, remain a question, which cannot be treated by the present approach. As seen from Fig. 5, the extended circuit model cannot explain a hump structure of $\sigma$ near 0.1 K.~\cite{rf:Bauch1,rf:Bauch2} 
We note that such a characteristic behavior has been observed in a DC-SQUID (superconducting quantum interference device) system composed only by low-$T_c$ superconductors.~\cite{rf:Balestro} While the deviation between our theory and the experimental data is left for a future problem, it might be attributed to the fact that our model neglects thermal/dynamical population from the quasi-ground state to the excited states in the metastable well. We note that in the nonadiabatic cases ($\omega_p \gg \omega_\mathrm{LC}$) corresponding to the MQT experiment for the YBCO biepitaxial junction~\cite{rf:Bauch1,rf:Bauch2} more quantum levels due to excitation of the LC circuit are relevant to the decay process than in a simple one-dimensional system. In order to investigate the thermal/dynamical population effect, we have to solve a master equation to obtain population probabilities of each level by the Larkin and Ovchinnikov theory.~\cite{rf:Larkin} This consideration may explain the anomaly of the escape rate $\Gamma(T)$ near the crossover temperature.

Finally we would like to comment advantages of the junction-LC coupling to the qubit and quantum optics applications.
The system that we considered in this paper can be regarded as an artificial atom (the Josephson junction) coupled to the quantized electromagnetic field (the LC circuit).
Therefore the appearance of several interesting phenomena relating to quantum optics,~\cite{rf:quantumoptics} e.g., the vacuum Rabi oscillation,~\cite{rf:vacuumRabi} generation of a non-classical state of the LC system,~\cite{rf:Schuster}  and laser oscillation~\cite{rf:NEClaser} is expected also in the YBCO biepitaxial junctions.
Additionally the LC circuit will act as a quantum information bus.~\cite{rf:Qubus,rf:Majer}
Therefore the entanglement or the coupling between separated high-$T_c$ qubits and eventually a high-$T_c$ version of the circuit-QED system~\cite{rf:cQED0,rf:cQED1,rf:cQED2} will be realized in such biepitaxial junctions.
These studies will open up the possibility of future applications for high-$T_c$ superconductor materials.

\acknowledgments
%%%%
%%%%
%%%%
%%%%
%%%%
We would like to thank J. Ankerhold, A. Barone, M. Fogelstr\"om, A. A. Golubov,  D. R. Gulevich, G. Johansson, J. R. Kirtley, T. L\"ofwander, F. Lombardi, J. P. Pekola, G. Rotoli, V. S. Shumeiko, and F. Tafuri  for useful discussions. 
One of the authors (S. K.) would like to thank the Applied Quantum Physics Laboratory at the Chalmers University of Technology for its hospitality during the course of this work.
This work was supported by Nano-NED Program under Project No. TCS.7029, CREST (JST), EU Nanoxide, EU STREP project MIDAS, the Swedish Research
Council (VR) under the Linnaeus Center on Engineered
Quantum Systems, the Swedish Research Council
(VR) under the project Macroscopic quantum tunneling and
coherence in superconductive d-wave junctions, the Knut
and Alice Wallenberg Foundation, the Swedish Foundation
for Strategic Research (SSF) under the project Oxide, and the JSPS-RSAS Scientist Exchange Program.

%%%%
%%%%
%%%%
%%%%
%%%%
\appendix
%%%%
%%%%
%%%%
%%%%
%%%%

%%%%%%%%%%
%%%%%%%%%%
%%%%%%%%%%
%%%%%%%%%%
\section{Finite Temperature Correction of the MQT escape rate}
%%%%%%%%%%
%%%%%%%%%%
%%%%%%%%%%
%%%%%%%%%%
If the temperature $T$ is much smaller than the crossover temperature $T_\mathrm{co}^\mathrm{2D}$, the finite-temperature MQT escape rate (in the weak retardation limit) is given by the product of $\Gamma_\mathrm{MQT}^\mathrm{2D} (T=0)$ and a finite-temperature correction as~\cite{rf:GrabertWeiss2}
\begin{eqnarray}
\Gamma_\mathrm{MQT}^\mathrm{2D} (T)
&=&
\Gamma_\mathrm{MQT}^\mathrm{2D} (T=0)
\exp \left[ {\cal A} (T)\right]
,
 \end{eqnarray}
where $ {\cal A} (T)$ is the finite-temperature correction to the bounce action and is given in term of the spectral density $J(\omega)$ as
\begin{eqnarray}
\!
{\cal A} (T)
\!
&=&
\!
\frac{\left( x_\mathrm{B} \tau_\mathrm{B} \right)^2}{2 \pi \hbar} 
\int_0^\infty d \omega J(\omega) 
\left[ \coth \left( \frac{\hbar \beta \omega}{2}\right) -1 \right]
.
 \end{eqnarray}
In  this equation $x_B \tau_B \equiv \int_{-\infty}^\infty d \tau x_B (\tau)$, where $x_B(\tau)$ is the zero temperature bounce.
For the spectral density Eq. (14), the enhancement function takes the form
\begin{eqnarray}
{\cal A} (T)
&=&
\frac{36 m \omega_\mathrm{LC}^3}{\hbar \omega_p^2}
(1- \gamma^2) 
\left[  
   \coth  \left( \frac{\hbar \beta \omega_\mathrm{LC}}{2}  \right) - 1
 \right] 
 \nonumber\\
 & \approx&
\frac{72 m \omega_\mathrm{LC}^3}{\hbar \omega_p^2}
(1- \gamma^2) 
e^{-\hbar \beta \omega_\mathrm{LC}}
,
 \end{eqnarray}
for $T \ll \hbar \omega_{LC} / 2 \pi k_B \equiv T_\mathrm{LC}$.
Thus the thermal enhancement  shows exponentially weak temperature dependence as long as $T \ll  T_\mathrm{LC}$.
On the other hand, for a damped system with $J(\omega) \propto \omega^s$, we get algebraic large enhancement ${\cal A} (T) \propto T^{1+s}$.~\cite{rf:GrabertWeiss2,rf:GrabertWeissHanggi}
Thus the coupling to the LC circuit gives the weak finite-temperature correction to the zero-temperature MQT escape rate $\Gamma_\mathrm{MQT}^\mathrm{2D} (T=0)$ for $T \ll T_\mathrm{LC}$.
Note that, in the actual YBCO junction,  $T_\mathrm{LC} < T_\mathrm{co}^\mathrm{2D}$.
In Sec. VI. C, we show the finite-temperature MQT escape rate $\Gamma_\mathrm{MQT}^\mathrm{2D} (T)$ for $0 < T < T_\mathrm{co}^\mathrm{2D}$.

%%%%%%%%%%
%%%%%%%%%%
%%%%%%%%%%
%%%%%%%%%%
\section{The escape rate in the crossover regime}
%%%%%%%%%%
%%%%%%%%%%
%%%%%%%%%%
%%%%%%%%%%
In this Appendix, we show an expression for the escape rate near $T_\mathrm{co}^\mathrm{2D}$.
The quantum enhancement factor $c_\mathrm{qm}^\mathrm{2D}$ in the TA escape rate $\Gamma_\mathrm{TA}^\mathrm{2D} (T)$ increases with decreasing the temperature,  and diverges as $T \to T_\mathrm{co}^\mathrm{2D}$.
This unphysical divergence is due to the appearance of a bounce trajectory for temperature below $T_\mathrm{co}^\mathrm{2D}$ and hence the breakdown of the Gaussian approximation.~\cite{rf:Weiss}
A crossover region is characterized by the condition $ \left| T - T_\mathrm{co}^\mathrm{2D}\right| \lesssim T_\mathrm{co}^\mathrm{2D}/\delta$ in which the dimensionless parameter $\delta$ is given by~\cite{rf:GrabertWeiss}
\begin{eqnarray}
\delta
&=&
\left[
 \omega_p^2 +
  \omega_R^2 \left( 1+ \frac{\partial \hat{\gamma} (\omega_R)}{\partial \omega_R}   \right)
 \right]
 \sqrt{
 \frac{M \beta}{2 B_4}
 }
.
 \end{eqnarray}
Here the coefficient $B_4$ measures the strength of the anharmonicity of the potential $U_\mathrm{1D}$ and is given by
\begin{eqnarray}
B_4
&=&
\omega_{p0}^2 
\left[
\frac{1}{ \sqrt{1-\gamma^2} }
- \frac{1}{2}
\frac{\omega_{p0}^2}{\omega_2^2 - \omega_p^2 +\omega_2 \hat{\gamma}(\omega_2)}
\right]
.
 \end{eqnarray}

In the crossover region, functional integral cannot be done by steepest descents, but requires a more careful treatment, leading to a escape rate of the form~\cite{rf:Hanggi}
\begin{eqnarray}
\Gamma_\mathrm{co}^\mathrm{2D} (T)
&=&
\frac{a}{2} 
\sqrt{
\frac{2 \pi}{\hbar | \tau'|}
}
\mathrm{erfc} \left[
\sqrt{
\frac{\hbar }{2| \tau'|}
}
\left(
\beta_\mathrm{co} -\beta
\right)
\right] 
\nonumber\\
&\times&
\exp
\left[
-\beta V_0^\mathrm{1D}
+
\frac{\hbar \left( \beta - \beta_\mathrm{co} \right)^2}{2 | \tau'|}
\right]
,
\end{eqnarray}
with $\mathrm{erfc}(x)=\int_{-\infty}^x d y \exp (-y^2/2)/\sqrt{2 \pi}$.
In this equation, $\beta_\mathrm{co}= 1/ k_B T_\mathrm{co}^\mathrm{2D}$, 
\begin{eqnarray}
a
&\equiv&
\frac{\omega_1^2 - \omega_p^2 + \omega_1 \hat{\gamma}(\omega_1)}
{ \omega_p^2 +
  \omega_R^2 \left( 1+ \frac{\partial \hat{\gamma} (\omega_R)}{\partial \omega_R}   \right)
}
c_\mathrm{qm}
,
\end{eqnarray}
 is the dimensionless prefactor, and $| \tau' | = \hbar \beta_\mathrm{co}^2 / 2 \kappa(\beta=\beta_\mathrm{co})^2$ is the energy derivative of the bounce period.
This expression applies for the temperature $T$ slightly above and below $T_\mathrm{co}^\mathrm{2D}$ and smoothly interpolates between the high-temperature escape rate $\Gamma_\mathrm{TA}^\mathrm{2D} (T)$ and the low-temperature escape rate  $\Gamma_\mathrm{MQT}^\mathrm{2D} (T)$ as demonstrated in Fig. 7.

%
%
%
%%
%%%
%%%
%%%
%===================================
\begin{figure}[t]
\begin{center}
\includegraphics[width=7cm]{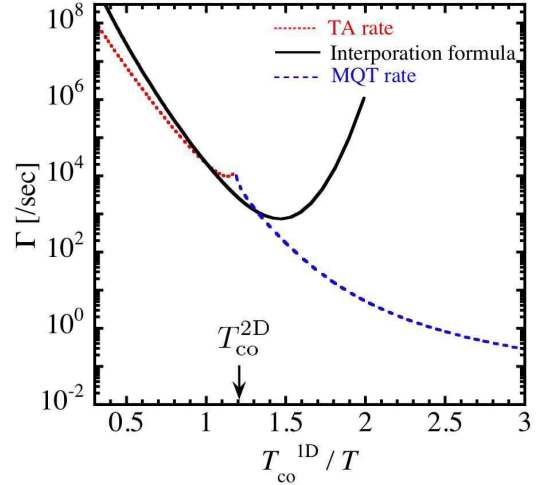}
\caption{(Color online) Temperature dependence of the escape rate near the crossover temperature $T_\mathrm{co}^\mathrm{2D}$. 
The TA escape rate $\Gamma_\mathrm{TA}^\mathrm{2D} (T)$ is represented by a dotted (red) line and the MQT escape rate $\Gamma_\mathrm{MQT}^\mathrm{2D} (T)$ as a dashed (blue) line.
The crossover formula $\Gamma_\mathrm{co}^\mathrm{2D} (T)$ (B3) is represented by a solid (black) line and smoothly matches onto these functions below and above the crossover temperature $T_\mathrm{co}^\mathrm{2D}$ (arrow).
Parameters are $I_C=1.4$ $\mu$A, $C_J=0.22$ pF, $L_S=$1.7 nH, and $C_S=1.6$ pF.}
\end{center}
\label{fig:7}
\end{figure}
%===================================
%%%
%%%
%%%
%%%%
 
\newpage

%%%%%%%%%%
%%%%%%%%%%
%%%%%%%%%%
%%%%%%%%%%

%
%
%
%

\begin{thebibliography}{99}
%%%%%%%%%%
%%%%%%%%%%
%%%%%%%%%%
%%%%%%%%%%
\bibitem{rf:MQT1}
S. Takagi, 
{\it Macroscopic Quantum Tunneling}  (Cambridge Univ. Press,  Cambridge, 2002).
%
%
\bibitem{rf:MQT2}
J. Ankerhold, 
{\it Quantum Tunneling in Complex Systems: The Semiclassical Approach}  (Springer-Verlag,  Berlin, 2007).
%
%
\bibitem{rf:Weiss}
U. Weiss, 
{\it Quantum Dissipative Systems}  (World Scientific, Singapore, 2008).
%
%
\bibitem{rf:MQT4}
E. Simanek, 
{\it Inhomogeneous Superconductors: Granular and Quantum Effects} (Oxford Univ. Press, 1994).
%
%
\bibitem{rf:Bauch1} T. Bauch, F. Lombardi, F. Tafuri, A. Barone, G. Rotoli, P. Delsing, and T. Claeson, 
Phys. Rev. Lett. {\bf 94}, 087003 (2005).
%
%
\bibitem{rf:Inomata1} K. Inomata, S. Sato, K. Nakajima, A. Tanaka, Y. Takano, H. B. Wang, M. Nagao, H. Hatano, and S. Kawabata, 
Phys. Rev. Lett. {\bf 95}, 107005 (2005).
%
%
\bibitem{rf:Jin} X. Y. Jin, J. Lisenfeld, Y. Koval, A. Lukashenko, A. V. Ustinov, and P. M\"uller, 
Phys. Rev. Lett. {\bf 96}, 177003 (2006).
%
%
\bibitem{rf:Matsumoto} T. Matsumoto, H. Kashiwaya, H. Shibata, S. Kashiwaya, S. Kawabata, H. Eisaki, Y. Yoshida, and Y. Tanaka, 
Supercond. Sci. Technol. {\bf 20}, S10 (2007).
%
%
\bibitem{rf:Li} S. X. Li, W. Qiu, S. Han, Y. F. Wei, X. B. Zhu, C. Z. Gu, S. P. Zhao, and H. B. Wang, 
Phys. Rev. Lett. {\bf 99}, 037002 (2007).
%
%
\bibitem{rf:Kashiwaya1} H. Kashiwaya, T. Matsumoto, H. Shibata, S. Kashiwaya, H. Eisaki, Y. Yoshida, S. Kawabata, and Y. Tanaka, 
J. Phys. Soc. Jpn. {\bf 77}, 104708 (2008).
%
%
\bibitem{rf:YurgensKadowaki} A. Yurgens, M. Torstensson, L. X. You, T. Bauch, D. Winkler, I. Kakeya, and K. Kadowaki,
Physica C {\bf 468}, 674 (2008).
%
%%
%
\bibitem{rf:Bauch2} T. Bauch, T. Lindstr\"om, F. Tafuri, G. Rotoli, P. Delsing, T. Claeson, and F. Lombardi, 
Science {\bf 311}, 57 (2006).
%
%
\bibitem{rf:Inomata2} K. Inomata, S. Sato, M. Kinjo, N. Kitabatake, H. B. Wang, T. Hatano, and K. Nakajima,
Supercond. Sci. Technol. {\bf 20}, S105 (2007). 
%
%
\bibitem{rf:Kashiwaya2} H. Kashiwaya, T. Matsumoto, S. Kashiwaya, H. Shibata, H. Eisaki, Y. Yoshida, S. Kawabata, and Y. Tanaka,
Physica C {\bf 468}, 1919 (2008).
%
%
\bibitem{rf:TafuriKIrtley} F. Tafuri and J. R. Kirtley,
Rep. Prog. Phys. {\bf 68}, 2573  (2005). 
%
%
\bibitem{rf:Yurgens} A. A. Yurgens, 
Supercond. Sci. Technol. {\bf 13}, R85 (2000). 
%
%
\bibitem{rf:Ioffe}
L B. Ioffe, V. B. Geshkenbein, M. V. Feigel'man, A. L. Fauch\'ere, and G. Blatter, 
Nature {\bf 398}, 679 (1999).
%
%
\bibitem{rf:Blais} 
A. Blais and A. M. Zagoskin, 
Phys. Rev. A {\bf 61}, 042308 (2000).
%
%
\bibitem{rf:Blatter} G. Blatter, V. B. Geshkenbein, and L. B. Ioffe, 
Phys. Rev. B {\bf 63}, 174511 (2001).
%
%
\bibitem{rf:Fominov}
Y. V. Fominov, A. A.  Golubov, and M. Kupriyanov,
JETP Lett. {\bf 77}, 587 (2003).
%
%
\bibitem{rf:Amin} 
M. H. S. Amin, A. Y. Smirnov, A. M. Zagoskin, T. Lindstr\"om, S. A. Charlebois, T. Claeson, and A. Y. Tzalenchuk, 
Phys. Rev. B {\bf 71}, 064516 (2005).
%
%
\bibitem{rf:Kato} T. Kato, A. A. Golubov, and Y. Nakamura, 
Phys. Rev. B {\bf 76}, 172502 (2007).
%
%
\bibitem{rf:Kawabata1} S. Kawabata, S. Kashiwaya, Y. Asano, and Y. Tanaka, 
Phys. Rev. B {\bf 70}, 132505 (2004). 
%
%
\bibitem{rf:Kawabata2} S. Kawabata, S. Kashiwaya, Y. Asano, and Y. Tanaka, 
Phys. Rev. B {\bf 72}, 052506 (2005). 
%
%
\bibitem{rf:Kawabata3} 
S. Kawabata, S. Kashiwaya, Y. Asano, and Y. Tanaka, 
Physica E  (Amsterdam) {\bf 29},  669 (2005).
%
%
\bibitem{rf:Kawabata4} S. Kawabata, S. Kashiwaya, Y. Asano, Y. Tanaka, T. Kato, and A. A. Golubov, 
Supercond. Sci. Technol. {\bf 20}, S6 (2007). 
%
%
\bibitem{rf:Kawabata5} S. Kawabata, A. A. Golubov, Ariando, C. J. M. Verwijs, H. Hilgenkamp, and J. R. Kirtley, 
Phys. Rev. B {\bf 76}, 064505 (2007). 
%
%
\bibitem{rf:Yokoyama} T. Yokoyama, S. Kawabata, T. Kato, and Y. Tanaka, 
Phys. Rev. B {\bf 76}, 134501 (2007). 
%
%
\bibitem{rf:Umeki} T. Umeki, T. Kato, T. Yokoyama, Y. Tanaka, S. Kawabata, and S. Kashiwaya, 
Physica C {\bf 463-465}, 157 (2007). 
%
%
\bibitem{rf:Tafuri} F. Tafuri, F. Carillo, F. Lombardi, F. Miletto Granozio, F. Ricci, U. Scotti di Uccio, A. Barone, G. Testa, E. Sarnelli, and J. R. Kirtley,
Phys. Rev. B {\bf 62}, 14431 (2000). 
%
%
\bibitem{rf:Lombardi0} F. Lombardi, F. Tafuri, F. Ricci, F. Miletto Granozio, A. Barone, G. Testa, E . Sarnelli, J. R. Kirtley, and C. C. Tsuei,
Phys. Rev. Lett. {\bf 89}, 207001 (2002). 
%
%
\bibitem{rf:Golubov} A. Golubov and F. Tafuri,
Phys. Rev. B {\bf 62}, 15200 (2000). 
%
%
\bibitem{rf:Rotoli}
G. Rotoli, T. Bauch, T. Lindstr\"om, D. Stornaiuolo,  F. Tafuri, and F. Lombardi,
Phys. Rev. B {\bf 75}, 144501 (2007).
%
%
\bibitem{rf:Lombardi}
F. Lombardi, T. Bauch, G. Rotoli, T. Lindstr\"om, J. Johansson, K. Cedergren, F. Tafuri, and T. Claeson,
IEEE Trans. Appl. Supercond. {\bf 17} 653 (2007).
%
%
\bibitem{rf:Zaikin}
G. Sch\"on and A. D. Zaikin,
Phys. Rep. {\bf 198}, 237 (1990).
%
%
\bibitem{rf:Chen}
Y. C. Chen, 
J. Low Temp. Phys. {\bf 65}, 133 (1986).
%
%
\bibitem{rf:Trees}
B. R. Trees, Y. H. Helal, J. S. Schiffrin, and B. M. Siller, 
Phys. Rev. B {\bf 76}, 224513 (2007).
%
%
\bibitem{rf:Leggett}
A. O. Caldeira and A. J. Leggett, 
Ann. Phys. (NY) {\bf 149}, 374 (1983).
%
%
\bibitem{rf:GrabertWeiss}
H. Grabert and U. Weiss,
Phys. Rev. Lett.  {\bf 53}, 1787 (1984).
%
%
\bibitem{rf:Fistul}
M. V. Fistul,
Phys. Rev. B  {\bf 75}, 014502 (2007).
%
%
\bibitem{rf:Wolynes}
P. G. Wolynes,
Phys. Rev. Lett.  {\bf 47}, 968 (1981).
%
%
\bibitem{rf:Esteve}
D. Esteve, M. H. Devoret, and J. M. Martinis,
Phys. Rev. B  {\bf 34}, 158 (1986).
%
%
\bibitem{rf:Martinis}
J. M. Martinis and H. Grabert,
Phys. Rev. B {\bf 38}, 2371 (1988).
%
%
\bibitem{rf:Zweger}
W. Zwerger,
Z. Phys. B {\bf 51}, 301 (1983).
%
%
\bibitem{rf:Yasui}
Y. Yasui, T. Takaai, and T. Ootsuka,
J. Phys. A: Math. Gen. {\bf 34}, 2643 (2001).
%
%
\bibitem{rf:Gerfand}
I. M. Gel'fand and A. M. Yaglom,
J. Math. Phys. {\bf 1}, 48 (1960).
%
%
\bibitem{rf:prefactor}
The retardation correction in the prefactor $A(T)$ gives negligible contribution to $\Gamma_\mathrm{MQT}^\mathrm{2D}$ in compared with that in the exponent ${\cal S}_T^\mathrm{2D} [x_B]$.
Therefore, in the calculation of $\Gamma_\mathrm{MQT}^\mathrm{2D}$, we have ignored the retardation correction in $A(T)$.
%
%
\bibitem{rf:Voss}
R. F. Voss and R. A. Webb, 
Phys. Rev. Lett.  {\bf 47}, 265 (1981).
%
%
\bibitem{rf:Garg} 
A. Garg, 
Phys. Rev. B{\bf 51}, 15592 (1995). 
%
%
\bibitem{rf:CJestimation} 
The junction capacitance $C_J$ was estimated by numerically calculating the electrostatic charge distribution in our biepitaxial junction geometry.   
%
%
%
%
\bibitem{rf:Sin2phi} 
 In the previous study,~\cite{rf:Bauch1} the remarkable temperature dependence of $\sigma(T)$ near $T=0.1$ K has been attributed to the anomalous temperature dependence  of  the higher harmonic component [$I_{C2} (T) $] of the Josephson current, i.e., $I_J = I_{C1} (T)  \sin \phi + I_{C2} (T) \sin 2 \phi$. However the measured temperature dependence of the total Josephson critical current  $I_{C1} (T) + I_{C2} (T) $ is almost constant up to 1 K (see inset of Fig. 3 in Ref. [\onlinecite{rf:Bauch1}]). In order to consistently explain these observations, the anomalous temperature dependence of $I_{C2} (T)$ should be completely compensated by $I_{C1} (T)$. Although opposite temperature-dependence between $I_{C1} (T)$ and $I_{C2}(T)$ has been observed experimentally in an in-plane $d$-wave junction and interpreted theoretically [E. Il'ichev et al. Phys. Rev. Lett. {\bf 86}, 5369 (2001)], the complete compensation between $I_{C1} (T)$ and $I_{C2} (T)$ for the $tilt$ $d$-wave junction is an open problem to be explained. We conjecture that the LC model adopted in this paper may provide more reasonable explanation for the anomalous hump structure. Detailed study is an interesting future problem. %
%
\bibitem{rf:Balestro}
F. Balestro, J. Claudon, J. P. Pekola, and O. Buisson, 
Phys. Rev. Lett.  {\bf 91}, 158301 (2003).
%
%
\bibitem{rf:Larkin} 
A. I. Larkin and Y. N. Ovchinnikov,
Sov. Phys. JETP {\bf 64}, 185 (1986).
%
%
%
%
\bibitem{rf:quantumoptics} 
L. Mandel and E. Wolf, 
{\it Optical Coherence and Quantum optics}  (Cambridge Univ. Press,  Cambridge, 1995).
%
%
\bibitem{rf:vacuumRabi} 
J. Johansson, S. Saito, T. Meno, H. Nakano, M. Ueda, K. Semba, and H. Takayanagi,
Phys. Rev. Lett.  {\bf 96}, 127006 (2006).
%
%
\bibitem{rf:Schuster} 
D. I. Schuster, A. A. Houck, J. A. Schreier, A. Wallraff, J. M. Gambetta, A. Blais, L. Frunzio, J. Majer, B. R. Johnson, M. H. Devoret, S. M. Girvin, and R. J. Schoelkopf,
Nature {\bf 445}, 515 (2007).
%
%
\bibitem{rf:NEClaser} 
O. Astafiev, K. Inomata, A. O. Niskanen, T. Yamamoto, Y. A. Pashkin, Y. Nakamura, and J. S. Tsai,
Nature {\bf 449}, 588 (2007).
%
%
\bibitem{rf:Qubus} 
T. P. Spiller, K. Nemoto, S. L. Braunstein, W. J. Munro, P. van Loock, and G. J. Milburn,
New J. Phys. {\bf 8}, 30 (2006). 
%
%
\bibitem{rf:Majer} 
J. Majer, J. M. Chow, J. M. Gambetta, J. Koch, B. R. Johnson, J. A. Schreier, L. Frunzio, D. I. Schuster, A. A. Houck, A. Wallraff, A. Blais, M. H. Devoret, S. M. Girvin, and R. J. Schoelkopf,
Nature  {\bf 449}, 443 (2007).
%
%
\bibitem{rf:cQED0} 
J. Q. You and F. Nori,
Phys. Rev. B  {\bf 68}, 064509 (2003).
%
%
\bibitem{rf:cQED1} 
A. Blais, R. -S. Huang, A. Wallraff, S. M. Girvin, and R. J. Schoelkopf,
Phys. Rev. A  {\bf 69}, 062320 (2004).
%
%
\bibitem{rf:cQED2}
A. Wallraff, D. I. Schuster, A. Blais, L. Frunzio, R. -S. Huang, J. Majer, S. Kumar, S. M. Girvin, and R. J. Schoelkopf,
Nature {\bf 431}, 162 (2004).
%
%
\bibitem{rf:GrabertWeiss2}
H. Grabert and U. Weiss,
Z. Phys. B  {\bf 56}, 171 (1984).
%
%
\bibitem{rf:GrabertWeissHanggi}
H. Grabert, U. Weiss, and P. H\"anggi,
Phys. Rev. Lett.  {\bf 52}, 2193 (1984).
%
%
\bibitem{rf:Hanggi}
P. H\"anggi and W. Hontscha,
J. Chem. Phys. {\bf 88}, 4094 (1988).
%
%
%
%
\end{thebibliography}
\end{document}